\documentstyle[aps,pra,eqsecnum,psfig,multicol,graphicx]{revtex}
\begin{document}
\draft
\title{Non-Markovian homodyne-mediated feedback on a two-level atom: a quantum
trajectory treatment}
\author{Jin Wang${}^{1}$, H.M. Wiseman${}^{2}$, G.J. Milburn${}^{3}$}
\address{${}^{1}$Centre for Laser Science,
Department of Physics, The University of Queensland, Brisbane,
Queensland 4072, Australia\\
${}^{2}$ School of Science, Griffith University, Brisbane, Queensland
4111, Australia \\
${}^{3}$Centre for Quantum Computer Technology,
The University of Queensland, Brisbane,
Queensland 4072, Australia}
\date{\today}
\maketitle

\begin{abstract}
Quantum feedback can stabilize a two-level
atom against decoherence (spontaneous emission), putting it into
an arbitrary (specified) pure state.  This requires
perfect homodyne detection of the atomic emission, and instantaneous
feedback. Inefficient detection was considered previously by two of us.
Here we allow for a non-zero
delay time $\tau$ in the feedback circuit. Because a two-level atom is a
nonlinear optical system, an analytical solution is not possible.
However, quantum trajectories allow a simple numerical simulation of
the resulting non-Markovian process. We find the effect of the time
delay to be qualitatively similar to that of inefficient detection.
The solution of the non-Markovian quantum trajectory will not remain
fixed, so that the time-averaged state will be mixed, not pure.
In the case where one tries to stabilize the atom in the excited state,
an approximate analytical solution to the quantum trajectory is
possible. The result, that the purity ($P=2{\rm Tr}[\rho^{2}]-1$)
of the average state is given by
$P=1-4\gamma\tau$ (where $\gamma$ is the
spontaneous emission rate)
is found to agree very well with the numerical results.
\end{abstract}


\newcommand{\beq}{\begin{equation}}
\newcommand{\eeq}{\end{equation}}
\newcommand{\bqa}{\begin{eqnarray}}
\newcommand{\eqa}{\end{eqnarray}}
\newcommand{\nn}{\nonumber}
\newcommand{\nl}[1]{\nn \\ && {#1}\,}
\newcommand{\erf}[1]{Eq.~(\ref{#1})}
\newcommand{\rf}[1]{(\ref{#1})}
\newcommand{\dg}{^\dagger}
\newcommand{\rt}[1]{\sqrt{#1}\,}
\newcommand{\smallfrac}[2]{\mbox{$\frac{#1}{#2}$}}
\newcommand{\half}{\smallfrac{1}{2}}
\newcommand{\bra}[1]{\langle{#1}|}
\newcommand{\ket}[1]{|{#1}\rangle}
\newcommand{\ip}[2]{\langle{#1}|{#2}\rangle}
\newcommand{\sch}{Schr\"odinger }
\newcommand{\schs}{Schr\"odinger's }
\newcommand{\hei}{Heisenberg }
\newcommand{\heis}{Heisenberg's }
\newcommand{\bl}{{\bigl(}}
\newcommand{\br}{{\bigr)}}
\newcommand{\ito}{It\^o }
\newcommand{\str}{Stratonovich }
\newcommand{\dbd}[1]{\frac{\partial}{\partial {#1}}}
\newcommand{\sq}[1]{\left[ {#1} \right]}
\newcommand{\cu}[1]{\left\{ {#1} \right\}}
\newcommand{\ro}[1]{\left( {#1} \right)}
\newcommand{\an}[1]{\left\langle{#1}\right\rangle}
\newcommand{\implies}{\Longrightarrow}

\begin{multicols}{2}

\section{Introduction}
The theory of open quantum systems is fundamental to understanding
dissipation on a microscopic and macroscopic level in diverse
fields (measurement theory, quantum optics, quantum chaos, solid
state, quantum computation) \cite{LD95,NG93,PG94,TP94,LV97},
wherever quantum irreversibility matters.
Traditionally, open quantum systems are described by the reduced density
operator,
which is obtained from the total density operator by tracing over the
environmental degrees of freedom. The dynamics of the reduced density
operator is
commonly described by a master equation.

Besides the density matrix formalism, there has been an increasing
interest over the last years in the quantum trajectory approach which uses a
stochastic Schr\"{o}dinger equation for the state
vector $\ket{\psi}$ \cite{Car93b,DalCasMol92,GarParZol92}.
This approach serves as a numerical tool for
solving the master equation,
as the reduced density operator is recovered as the ensemble average of
these stochastically evolving pure states.
It also has a deeper significance, as, in some cases, the
quantum trajectories can be interpreted as the evolution of the system
state conditioned on continuous monitoring of its environment
\cite{WisMil93c,Wis96a}.
This means that quantum trajectories are ideal for treating quantum
feedback, where the measurement results from this monitoring are used
to control the dynamics of the system \cite{WisMil93b,Wis94a}.

The history of feedback-control in open quantum systems
 goes back to 1980's with the work
of Yamamoto and co-workers \cite{HauYam86,YamImoMac86},
and Shapiro and co-workers \cite{Sha87}.
Their objective was to explain the observation \cite{WalJak85a} of
sub-shot-noise fluctuations in an
in-loop photocurrent. They did this using quantum Langevin equations
(stochastic Heisenberg equations for the system operators) and also
semiclassical techniques. The latter approach was made fully
quantum-mechanical by Plimak \cite{Pli94}. For systems with linear
dynamics, all of these approaches, and the quantum trajectory approach of
Refs.~\cite{WisMil93b,Wis94a}, are equally easy to use to find
analytical solutions. The advantage of the Wiseman and Milburn
(quantum trajectory) approach to quantum control via feedback
is for systems with nonlinear dynamics, as we will discuss
later.

Before proceeding further,
it is necessary to distinguish what we mean by `quantum
control' via feedback from an alternative usage of the
term `quantum control' which has become current in laser
chemistry \cite{Rabitz}.
The work in this field has sought sufficient control over laser pulses so
that very specific quantum dynamics of electronic and vibrational
states may be
achieved, primarily with the objective of making and
breaking bonds in molecular
systems. Typically such experiments are conducted on a large ensemble
of identical prepared
target systems. Two kinds of control
are distinguished; (i) learning control,
and, (ii) feedback control. In learning control the objective is to design
the required laser pulse as follows.
After applying a pulse to a sample of many identical constituents (an
ensemble),
one measures some property of the system, adjusts
the laser pulse according to the measurement
results, then applies the new laser pulse to a new sample and so on.
The cycle of adjust-excite-measure
is repeated until the desired results are obtained.  In feedback control,
the laser excitation always acts on the {\em same} sample
of many identical systems
without re-preparation, through the same repeated
sequence. Of course the adjustment one makes to the next pulse will
be quite different in the  two cases.

In contrast to this controlling of the average state of an ensemble,
 we are concerned with the control of the state of a {\em single} quantum
system
subject to continuous measurement.
While this is similar to (ii), the fact that
only a single quantum system is involved, not an ensemble, makes it
fundamentally different.
The quantum efficiency of the  measurement has to be
relatively large for the control to be at all effective. This
means that one has to take into account the quantum back-action of the
measurement on the system, which is not necessary in the laser
chemistry `feedback control' referred to above. Only very
recently has it become possible to isolate, and
continuously interrogate a single quantum
system at the back-action limit. This has been achieved in a
number of fields including cavity QED \cite{Rai94}
ion trapping \cite{NIST} and mesoscopic
electronics \cite{Buks}.

Returning to the different approaches to control of a single quantum
system, another advantage of the Wiseman-Milburn quantum feedback theory is
that it is very easy to consider the limit of Markovian (i.e.
instantaneous) feedback. In this limit it is possible to derive a
master equation which describes the unconditioned
system dynamics including the effect of
feedback \cite{WisMil93b,Wis94a}. In previous work, two of us
\cite{WangWise} applied Markovian quantum feedback theory to a
two-level atom, where the light emitted by the atom was detected
 and used to control the dynamics of the atom.
Using the feedback master equation, analytical results are easily
derived. We showed that, with one exception, the feedback can generate an
arbitrary
stable pure state of the atom. This is an example of using feedback
to control decoherence (due to
spontaneous emission).

Decoherence in open quantum  systems refers to the tendency for pure system
states to become entangled with many different states of
the environment \cite{Zurek}. If the
states of the environment are then averaged out, the system state
tends to become mixed. The radiative decay of a two level atom with a
non-zero dipole is an example. In this case the atomic state becomes
correlated
with the electromagnetic field states.

Our previous work \cite{WangWise} was motivated by the 1998 work of Hofmann,
Mahler, and Hess (HMH)
\cite{HofHesMah98}, in which it was  shown that
by making part of the coherent driving of a two-level atom
proportional to the homodyne
photocurrent, it was possible to stabilize the state at any point on the
bottom half of the Bloch sphere.
Our work \cite{WangWise}, using the
Wiseman-Milburn feedback theory, reanalyzed their proposal
and generalized their results in two ways. First, we showed that
any point
on the  upper or lower half (but not the equator) of the Bloch sphere may
be stabilized. Second, we
considered non-unit-efficiency detection, and quantified the effectiveness
of the feedback by calculating the
maximal purity obtainable in any particular direction in Bloch space.

It is now of interest to extend our theory of feedback-mediated decoherence
control to a
non-Markovian process. This would arise naturally in an experiment,
since any feedback apparatus would necessarily have some delay
associated with it. It is also of interest theoretically because
non-Markovian feedback in a nonlinear system has not been considered
in detail before. The two-level atom, being the simplest possible
system with nonlinear dynamics, is an ideal testing ground for
descriptions of non-Markovian processes. Of course the master
equations derived in Refs.~\cite{WisMil93b,Wis94a} cannot be applied to
non-Markovian feedback in which the delay is significant on the time scale
of the
system.

Most of previous approaches to non-Markovian feedback have been limited to
linear systems such as an optical cavity mode with feedback based on
homodyne-detection. Wiseman and Milburn \cite{WisMil94a} used a quantum
trajectory treatment, as did Doherty and Jacobs \cite{DohJac99}.
Giovannetti, Tombesi and  Vitali \cite {VGI99}
used the quantum Langevin approach of input-output theory.
As noted above, linear systems allow analytical solutions, and all
approaches to quantum feedback work equally as well. However, for a
nonlinear system like a two-level atom an analytical solution for
non-Markovian feedback is not possible in general. Hence a numerical solution
is necessary. The only practical way of doing such a
simulation is to use the stochastic quantum trajectories for the
conditioned system state, which underly the Wiseman-Milburn feedback
theory. Such a numerical quantum trajectory simulation
 was done recently for a non-linear system consisting  of a
particle in a double-well potential with feedback determined by LQG
(linearized
quadratic Gaussian) control theory \cite{Andrew2000}.

Quantum trajectories describing non-Markovian feedback are of some
intrinsic interest as they are a special example of non-Markovian
quantum trajectories. These have been the subject of much interest
 lately. An example of this is the stochastic wave function method developed by
Jack, Collett and Walls \cite{walls2000}, where the non-Markovian
stochastic equation of motion for the state vector involves a multiple time
integration
over the system's history conditioned on the measurement record over a
finite time interval.
Furthermore, it has been shown recently by Diosi and Strunz and
co-workers
\cite {Laj97,WTS99} that
it is in principle
possible to  construct a stochastic Schr\"{o}dinger equation which
describes the non-Markovian time
evolution of any open quantum system.

Using non-Markovian quantum trajectory simulations we are able to
determine the effect of a time delay on the feedback stabilization
scheme introduced in Refs.~\cite{HofHesMah98,WangWise}. We find that the
effect of a time delay is qualitatively similar to that of inefficient
detectors. That is, the average state can no longer be made pure.
States in the upper half of the Bloch sphere are more affected than
those in the lower half, and states near the equator are affected most
of all. In one special case, where the feedback aims to stabilize the
atom in the excited state, it is possible to linearize the quantum
trajectory if the time delay is short,
and so obtain an approximate analytical solution. The
result compares very well with the numerical solution. It is not
obvious that this linearization could be performed using any of
the other approaches to quantum feedback.

The paper is organized as following. In Sec.~II we introduce the
measurement scheme (homodyne detection) for our system (the two-level
atom). In Sec.~III, we briefly review our previous work of
Markovian feedback stabilization.
The effectiveness of non-Markovian feedback is studied in Sec.~IV,
both numerically and analytically.
In Sec.~V we conclude with a discussion of our results.

\section{The Measurement Scheme}

\subsection{The driven, damped atom}

\begin{figure}[tbp]
\includegraphics[width=0.45\textwidth]{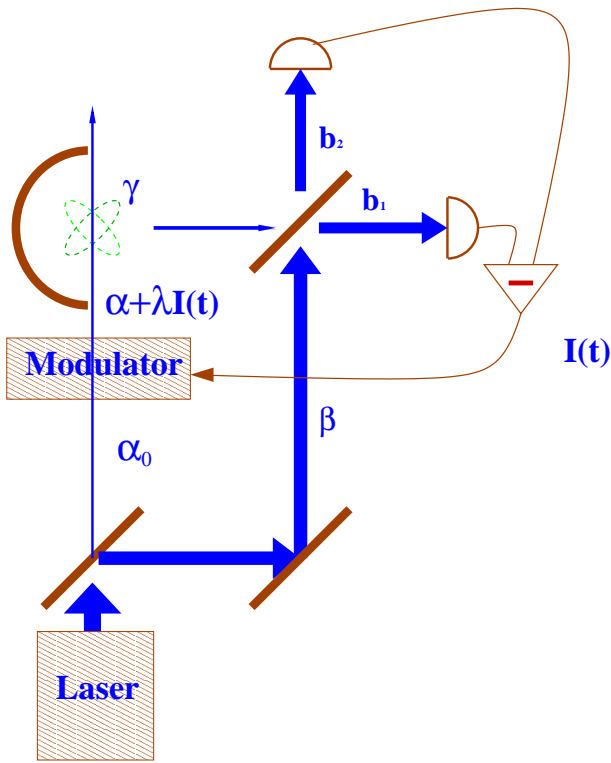}
\caption{\narrowtext Diagram of the experimental apparatus. The laser
beam is split to produce both the local oscillator $\beta$
and the field $\alpha_{0}$ which is modulated using the current $I(t)$.
The modulated beam,
with amplitude proportional to $\alpha+\lambda I(t)$, drives an atom
at the center of the parabolic mirror. The fluorescence thus
collected is subject to homodyne detection using the local
oscillator, and gives rise to the homodyne photocurrent $I(t)$.}
	\protect\label{fig:diag}
\end{figure}

Consider an atom, with two relevant levels
$\{\ket{g},\ket{e}\}$ and lowering operator
$\sigma=\ket{g}\bra{e}$. Let the  decay rate be $\gamma$, and let
it be driven by a resonant classical driving field with Rabi
frequency $2\alpha$. This is as shown in Fig.~\ref{fig:diag},
where for the moment
we are omitting feedback by setting $\lambda=0$.
This system is well-approximated by the master
equation
\beq \label{me1}
\dot{\rho} = \gamma{\cal D}[\sigma]\rho - i\alpha [\sigma_{y},\rho],
\eeq
where the Lindblad \cite{Lin76} superoperator is defined as usual
${\cal D}[A]B \equiv ABA\dg - \{A\dg A,B\}/2$.
In this master equation
 we have chosen to define the $\sigma_{x}=\sigma+\sigma\dg$ and
$\sigma_{y}=i\sigma-i\sigma\dg$ quadratures of the atomic dipole
relative to the driving field. The effect of driving is to rotate the
atom in Bloch space around the $y$-axis. The state of the atom in
Bloch space is described by the three-vector $(x,y,z)$. It is related
to the state matrix $\rho$ by
\beq \label{Bloch}
\rho = \frac{1}{2}\ro{I + x\sigma_{x}+y \sigma_{y}+z\sigma_{z}}.
\eeq

It is easy to show that the stationary solution of the master
equation (\ref{me1}) is
\bqa
{x}_{\rm ss}&=&\frac{-4\alpha\gamma}{\gamma^2+8\alpha^2},\\
{y}_{\rm ss}&=&0         ,\\
{z}_{\rm ss}&=&\frac{-\gamma^2}{\gamma^2+8\alpha^2}.
\eqa
For $\gamma$ fixed, this is a family of solutions parameterized by the
driving strength $\alpha \in (-\infty,\infty)$. All members of the
family are in the $x$--$z$ plane on the Bloch sphere. Thus for this
purpose we can reparametrize the relevant states using $r$ and
$\theta$ by
\beq\label{rtxz}
x = r\sin \theta, \;\;\;
z =r\cos \theta, \label{rtz}
\eeq
where $\theta \in [-\pi,\pi]$. Since
\beq
{\rm Tr}[\rho^{2}] = \frac{1}{2}\ro{1 + x^{2}+y^{2}+z^{2}}
\eeq
is a measure of the purity of the Bloch sphere, $r=\sqrt{x^{2}+z^{2}}$,
the distance from
the center of the sphere, is also a measure
of purity. Pure states correspond to $r=1$ and maximally mixed
states to $r=0$.

\begin{figure}[tbp]
\includegraphics[width=0.45\textwidth]{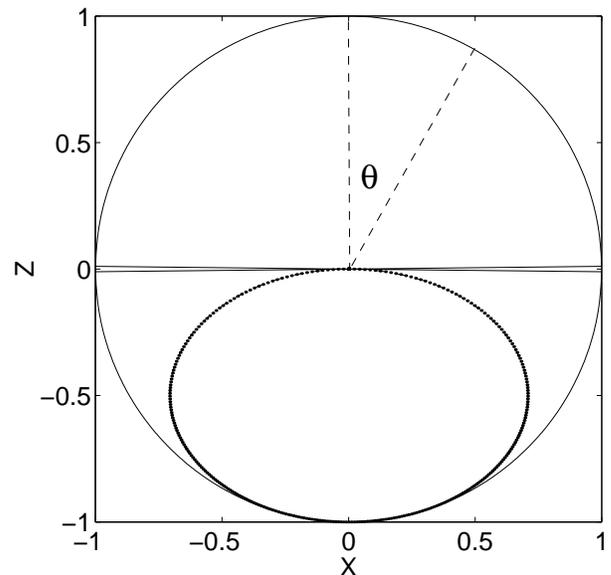}
\caption{\narrowtext Locus of the solutions to the Bloch equations.
The ellipse in the lower half plane is the locus for the equations
with driving only. The full circle (minus the points on the equator)
is the locus for the equations with optimal driving and feedback, as
defined in Sec.~III.}
	\protect\label{fig:bs1}
\end{figure}

The locus of
solutions in this plane (an ellipse) is shown in Fig.~\ref{fig:bs1}. Since
$z_{\rm ss}<0$, all solutions are in the lower half of the Bloch sphere.
That is, we are restricted to $|\theta| > \pi/2$. Also, it is evident
that the smaller $|\theta|$ is (that is, the more excited the atom is),
the smaller $r$ is (that is, the less pure the atom is). At
$|\theta|=\pi$ the stationary state is pure, but this is not
surprising as it is simply the ground state of the atom with no
driving. As
$|\theta|\to \pi/2$ we have $r \to 0$. This can only be approached
asymptotically as $|\alpha| \to \infty$.
In summary, the stationary states we can reach by driving
the atom are limited, and generally far from pure.

\subsection{Homodyne detection}
Now consider subjecting the atom to homodyne detection. As shown in
Fig.~\ref{fig:diag}, we assume that all of the fluorescence of the
atom is collected and turned into a beam (represented in
Fig.~\ref{fig:diag} by
placing the atom at the focus of a mirror). Ignoring the vacuum
fluctuations in the field, the annihilation operator for this beam is
$\sqrt{\gamma}\sigma$, normalized so that the mean intensity
$\gamma\an{\sigma\dg \sigma}$ is equal to
the number of photons per unit time in the beam.
This beam then enters one
port of a 50:50 beam splitter, while a strong local oscillator
$\beta$ enters the other. To ensure that this local oscillator has a
fixed phase relationship with the driving laser used in the measurement, it
would be
natural to utilize the same coherent light field source in the
driving process
and as the local oscillator in the homodyne detection. This is as
shown in Fig.~\ref{fig:diag}.

Again ignoring vacuum fluctuations, the two field operators exiting
the beam splitter, $b_{1}$ and $b_{2}$, are
\beq
b_{k} = \sq{\sqrt{\gamma}\sigma -(-1)^{k}\beta}/\sqrt{2}.
\eeq
When these two fields are detected, the two photocurrents produced
have means
\beq \label{meanI}
\bar{I}_{k} = \an{ |\beta|^{2} - (-1)^{k}\ro{ \sqrt{\gamma}\beta\sigma\dg +
\sqrt{\gamma}\sigma \beta^{*}} + \gamma\sigma\dg\sigma}/2.
\eeq
The middle two terms represent the interference
between the system and the local
oscillator.

Equation \erf{meanI} gives only the mean photocurrent. In an
individual run of the experiment for a system, what is recorded is not
the mean photocurrent, but the instantaneous photocurrent. This
photocurrent will vary stochastically from one run to the next,
because of the irreducible randomness in the quantum measurement
process. This randomness is not just noise, however. It is correlated
with the evolution of the system and thus tells the experimenter
something about the state of the system. In fact, if the detection
efficiency is perfect, the system is collapsed into a pure state,
rather than the mixed state which is the solution of the master
equation. The stochastic evolution of the state of the system
conditioned on the measurement record is called a ``quantum
trajectory'' \cite{Car93b}. Of course, the master equation is still
obeyed on average, so the set of possible quantum trajectories is
called an unraveling of the master equation \cite{Car93b}.
It is the conditioning of the system state on the photocurrent record
that allows feedback of the photocurrent to control the system
state.  The
application of an appropriate feedback loop to this continuous measurement
process (to be considered in Sec.~III)
realizes an effective ``reservoir engineering'' to control the
system at the quantum level.

The ideal limit of homodyne detection is when the local
oscillator amplitude
goes to infinity, which in practical terms means
$|\beta|^2\gg \gamma $. In this limit,
the rate of the photodetections goes
to infinity and thus it should be possible to change the point process of
photocounts into a continuous photocurrent with white noise. Also,
the only relevant quantity is the difference between the two photocurrents.
Suitably normalized, this is \cite{Car93b,WisMil93a}
\beq \label{homo1}
I(t) = \frac{I_{1}(t)-I_{2}(t)}{|\beta|} =
\sqrt{\gamma}\an{e^{-i\Phi}\sigma\dg + e^{i\Phi}\sigma}_{\rm c}(t) + \xi(t).
\eeq

A number of aspects of \erf{homo1} need to be explained. First,
$\Phi = \arg\beta$, the phase of the local oscillator (defined relative to the
driving field). Second, the subscript c means conditioned and refers
to the fact that if one is making a homodyne measurement then this
yields information about the system. Hence, any system averages will
be conditioned on the previous photocurrent record. Third, the final
term $\xi(t)$ represents Gaussian white noise, so that
\beq
\xi(t)dt = dW(t),
\eeq
an infinitesimal Wiener increment defined by \cite{Gar85}
\bqa
dW(t)^2=dt , \label{ito1}\\
{\rm E}[dW(t)]=0 . \label{dW0}
\eqa

Since the stationary solution of the master equation
confines the state to the $x$--$z$ plane, it makes sense to follow HMH
by setting $\Phi=0$. In that case,
\beq \label{homo2}
I(t) =\sqrt{\gamma}\an{\sigma_{x}}_{\rm c}(t) + \xi(t).
\eeq
That is,
the deterministic part of the
homodyne photocurrent is proportional to $x_{\rm c} =
\an{\sigma_{x}}_{\rm c}$. This should be useful for controlling
the dynamics of the state in the $x$--$z$ plane by feedback, as we
will consider in Sec.~III. Of course, all that really matters here is
the relationship between the driving phase and the local oscillator
phase, not the absolute phase of either.

The conditioning process referred to above can be made explicit by
calculating how the system state changes in response to the measured
photocurrent. Assuming that the state at some point in time is pure
(which will tend to happen because of the conditioning anyway), its
future evolution can be described by the stochastic \sch equation SSE.
\cite{Car93b,WisMil93a}
\beq \label{SSE1}
d\ket{\psi_{\rm c}(t)} = \hat{A}_{\rm c}(t)\ket{\psi_{\rm c}(t)}dt
+ \hat{B}_{\rm c}(t)\ket{\psi_{\rm c}(t)}dW(t).
\eeq
This is an \ito stochastic equation \cite{Gar85}
with a drift term and a diffusion
term. The operator for the drift term is
\beq
\hat{A}_{\rm c}(t) = \frac{\gamma}{2}\sq{-\sigma\dg\sigma
+ \an{\sigma_{x}}_{\rm c}(t)\sigma - \an{\sigma_{x}}_{\rm c}^{2}(t)/4}
-i\alpha\sigma_{y},
\eeq
while that for the diffusion is
\beq
\hat{B}_{\rm c}(t) = \sqrt{\gamma}\sq{\sigma - \an{\sigma_{x}}_{\rm
c}(t)/2}.
\eeq
Both of these operators are conditioned in that they depend on the
system average
\beq
\an{\sigma_{x}}_{\rm c}(t) = \bra{\psi_{\rm c}(t)}\sigma_{x}
\ket{\psi_{\rm c}(t)}.
\eeq

As stated above, on average
 the system still obeys the master equation (\ref{me1}).
This is easiest to see from the stochastic master equation
(SME), which allows for impure initial conditions. The SME can be
derived from the SSE by constructing
\bqa
d \ro{\ket{\psi_{\rm c}}\bra{\psi_{\rm c}}} &=&
\ro{d\ket{\psi_{\rm c}}}\bra{\psi_{\rm c}} +
\ket{\psi_{\rm c}}\ro{d\bra{\psi_{\rm c}}} \nl{+}
\ro{d\ket{\psi_{\rm c}}}\ro{d\bra{\psi_{\rm c}}}, \label{SSE2SME}
\eqa
using the \ito rule (\ref{ito1}),
and then identifying $\ket{\psi_{\rm c}}\bra{\psi_{\rm c}}$ with
$\rho_{\rm c}$. The result is
\beq \label{SME1}
d\rho_{\rm c} = dt\gamma{\cal D}[\sigma]\rho_{\rm c} - idt\alpha
[\sigma_{y},\rho_{\rm c}] +
dW(t)\sqrt{\gamma}{\cal H}[\sigma]\rho_{\rm c} ,
\eeq
where ${\cal H}[A]B \equiv AB+BA\dg - {\rm Tr}[AB+BA\dg]B$. Although
this has been derived assuming pure initial conditions, it is valid
for any initial conditions \cite{WisMil93a}. This is
also an \ito equation, which means the evolution for the
ensemble average state matrix
\beq
\rho(t)={\rm E}[\rho_{\rm c}(t)]
\eeq
is found simply by averaging over the photocurrent
noise term by using \erf{dW0}.
This procedure yields the original master equation (\ref{me1}) again.
The term ``quantum trajectory'' can be applied to any
 stochastic conditioned evolution of the
system, be it described by a SSE or SME.

\section{Markovian feedback Stabilization}
We now include feedback onto the amplitude of the driving on the atom,
proportional to the homodyne photocurrent, as done by HMH.
This is as shown in Fig.~\ref{fig:diag}, where the driving field
passes through an electro-optic
amplitude modulator controlled by the photocurrent, yielding a field
proportional to $\alpha + \lambda I(t)$. This means that the feedback
can be described by the Hamiltonian
\beq \label{fbH}
H_{\rm fb} = \lambda \sigma_{y}I(t).
\eeq
In this section we are assuming instantaneous feedback, while feedback
with time
delay will be considered in the next section.

Since the homodyne photocurrent (\ref{homo1}) is defined in terms of
system averages and the noise $dW(t)$, the SSE including feedback can
still be written as an equation of the form (\ref{SSE1}). The
effect of the
feedback Hamiltonian can be shown \cite{WisMil93b,WisMil94a} to
change the drift and diffusion operators to
\bqa
\hat{A}_{\rm c}(t) &=& \frac{\gamma}{2}\sq{-\sigma\dg\sigma
+ \an{\sigma_{x}}_{\rm c}(t)\sigma - \an{\sigma_{x}}_{\rm c}^{2}(t)/4}
-i\alpha\sigma_{y}
\nl{+} \lambda \sqrt{\gamma}
\sq{-i \an{\sigma_{x}}_{\rm c}(t)\sigma_{y} -2\sigma\dg \sigma}  -
{\lambda^{2}}/{2}, \\
\hat{B}_{\rm c}(t) &=& \sqrt{\gamma}\sq{\sigma - \an{\sigma_{x}}_{\rm
c}(t)/2} - i\lambda\sigma_{y}.
\eqa

Say we wish to stabilize the pure state with Bloch angle $\theta_0$,
as defined in Eq.~(\ref{rtxz}), with $r=1$ of course.
In terms of the ground and excited states, this state is
\beq
 \ket{\theta_0}=\cos\frac{\theta_0}{2}\ket{e}+\sin\frac{\theta_0}{2}\ket{g}.
\eeq
Now for this state to be stabilized we must have
\beq
\sq{\hat{A}_{\rm c}(t)dt + \hat{B}_{\rm c}(t)dW(t)}\ket{\theta_0}
\propto \ket{\theta_0}.
\eeq
We cannot say the left-hand-side should equal zero because a change in
the overall phase still leaves the physical state unchanged. However,
we can work with this equation, and simplify it by dropping all terms
proportional to the identity operator in $\hat{A}_{\rm c}(t)$ and
$\hat{B}_{\rm c}(t)$. We can also demand that it be satisfied for the
deterministic and noise terms separately, because $dW(t)$ can take any
value. This gives the two equations
\bqa
\ro{ \sqrt{\gamma}\sigma -i\lambda\sigma_{y}}\ket{\theta_0} &\propto&
\ket{\theta_0} ,\\
\left[\gamma\ro{-\sigma\dg\sigma
+ \sin\theta_0 \sigma}
-i2\alpha\sigma_{y}\phantom{\ket{\sqrt{\gamma}}}\right. && \nn \\
\left. +\, \lambda \sqrt{\gamma}
\ro{-i \sin\theta_0\sigma_{y}/2 -\sigma\dg \sigma}\right] \ket{\theta_0}
&\propto& \ket{\theta_0},
\eqa
where we have put $\an{\sigma_{x}}_{\rm c}(t)$ equal to $\sin\theta_0$,
its value for the state $\ket{\theta_0}$.

Solving the first equation easily yields the condition
\beq \label{lth}
\lambda = -\frac{\sqrt{\gamma}}{2}(1+\cos\theta_0).
\eeq
Substituting this into the second equation gives, after some
trigonometric manipulation, the second condition
\beq \label{ath}
\alpha = \frac{\gamma}{4}\sin\theta_{0}\cos\theta_0.
\eeq
These set of driving and feedback amplitude can stabilize the system in any
particular state in the Bloch sphere (except those on the equator; see
below) for unit-detection efficiency and
Markovian feedback. These features are
illustrated in Fig.~2 as a full circle (minus the points on the
equator)
which is the locus for
the equations
with optimal driving and feedback. This is contrary to the conclusion of HMH
\cite{HofHesMah98},
based on a linearized stability analysis, that ``long term stability
of \ldots inverted states [i.e. states in the upper half plane]
cannot be achieved.''

Once again, it is convenient to construct the stochastic master
equation (SME). This can be derived from the SSE in the same way as
before
[\erf{SSE2SME}]. The
result is \cite{WisMil93b,WisMil94a}
\bqa
d\rho_{\rm c} &=&  dt\gamma{\cal D}[\sigma]\rho_{\rm c} - idt\alpha
[\sigma_{y},\rho_{\rm c}] \nl{-}
idt \lambda [\sigma_{y},\sigma\rho_{\rm c} + \rho_{\rm c}\sigma\dg] +
dt(\lambda^{2}/\gamma){\cal D}[\sigma_{y}]\rho_{\rm c} \nl{+}
dW(t){\cal H}[\sqrt{\gamma}\sigma-i\lambda\sigma_{y}]\rho_{\rm c}.
\eqa
Also as before, this is an \ito stochastic equation, which means that
the ensemble average can be found simply by dropping the stochastic
terms. This time, the result is not the original master equation, but
rather the feedback-modified master equation
\beq
\dot{\rho} = -i[\alpha\sigma_{y},\rho]+{\cal
D}[\sqrt{\gamma}\sigma-i\lambda\sigma_{y}]\rho \equiv {\cal L}\rho.
\eeq
Here we have put the Liouvillian superoperator ${\cal L}$ in
 a manifestly Lindblad form.

 It can be shown that  the pure state $\rho =
\ket{\theta_{0}}\bra{\theta_{0}}$ is a stationary solution of this
master equation, as expected. However, if and only if
 $\theta_{0}=\pm\pi/2$, it is not a
stable solution. That is because the master equation for
$\theta_{0}=\pm\pi/2$ (it is the same for both cases) has more than one
null eigenvalue, and any mixture of the two equatorial states will be
a stationary solution. This means that states on the equator,
which are equal superpositions of excited and ground
states, cannot be
well-protected against decoherence. We will return to this in Sec.~IV~C.

\section{Non-Markovian Feedback}
\subsection{Non-Markovian feedback SME}
Now we consider the case of non-Markovian feedback. This would
occur experimentally unless the response of the feedback apparatus is flat in
frequency space over a bandwidth much larger
than any relevant system rate. For simplicity, we consider the case of
just having a time delay $\tau$. That is, we take the feedback
Hamiltonian to be
\beq \label{fbH2}
H_{\rm fb} = \lambda \sigma_{y}I(t-\tau).
\eeq
The Stochastic Master Equation (SME ) of the total conditioned evolution of
the system
for $\tau$ finite is given by \cite{WisMil93b,Wis94a}
\bqa
d\rho_{\rm c} &=&  dt\gamma{\cal D}[\sigma]\rho_{\rm c} - idt\alpha
[\sigma_{y},\rho_{\rm c}] + dW(t)\sqrt{\gamma}{\cal H}[\sigma]\rho_{\rm
c}\nl{-}
\lambda dt I(t-\tau)i[\sigma_{y},\rho_{\rm c}] - dt(\lambda^{2}/2)
[\sigma_{y},[\sigma_{y},\rho_{\rm c}]] .
 \label{SME2}
\eqa
Here $I(s)$ is still given by \erf{homo2}, so that the noise $dW(s)$
appears twice in \erf{SME2}, with two different time arguments, $s=t$
and $s=t-\tau$.

In \erf{SME2} it is no longer possible simply to set the noise $dW$ equal
to its
expectation value of zero to obtain a deterministic equation. That is
because the $dW(t-\tau)$ appearing in the feedback through $I(t-\tau)$
is correlated with the system state, since it already conditioned it
through the measurement term at time $t-\tau$. Also, the atomic
dynamics are nonlinear, so there is no option but to solve \erf{SME2}
numerically.
The most convenient way to treat the stochastic dynamics is, in general, to use
stochastic Bloch Equations (SBE). These are simply the
stochastic equations for the conditioned Bloch vector, defined by
\beq
\rho_{\rm c} =  \frac{1}{2}\ro{I + x_{\rm c}\sigma_{x} +
y_{\rm c} \sigma_{y} + z_{\rm c}\sigma_{z}}.
\eeq
The SBE corresponding to \erf{SME2} are
\bqa
\left(\begin{array}{c}
dx_{\rm c}\\dy_{\rm c}\\dz_{\rm c}\end{array}\right)
&=&\left(\begin{array}{ccc}
-\gamma/2-2{\lambda^2}&0  &  2{\alpha}+2\lambda{I_{-\tau}}\\
0   & -{\gamma}/{2} & 0       \\
-2\alpha-2\lambda{I_{-\tau}}  & 0 &-\gamma-2{\lambda}^2\\
\end{array}\right)\nl{\times}
\left(\begin{array}{c}
x_{\rm c}\\y_{\rm c}\\z_{\rm c}\end{array}\right)dt
-dt\left(\begin{array}{c}
0\\0\\-\gamma\end{array}\right) \nl{+}
 \sqrt{\gamma}\, dW(t)\left(\begin{array}{c}
1-x_{\rm c}^2+z_{\rm c}\\
-x_{\rm c}y_{\rm c}\\
-x_{\rm c}-x_{\rm c}z_{\rm c}\end{array}\right), \label{SBE1}
\eqa
Here we have used $I_{-\tau}$ as short-hand for $I(t-\tau)$.

We can use these equations to perform the numerical simulation of the
non-Markovian process. In the numerical simulation, we have not optimized
the driving and
feedback amplitude to achieve the best possible result for a
given time delay $\tau$. This is as opposed to the Markovian case
with inefficient detection where we optimized for every $\eta$.
The difference is because in the present case the optimization
problem is considerable more difficult.
 However, since we expect that the feedback mechanism will only work
effectively
with a small feedback delay time, it is thus convenient and reasonable to use
the same set of
driving and feedback amplitude as in the Markovian case, namely
\bqa \label{par1}
\lambda &=& -\frac{\sqrt{\gamma}}{2}(1+\cos\theta_0),\\
\alpha &=& \frac{\gamma}{4}\sin\theta_0\cos\theta_0.
\label{par2}
\eqa

\subsection{The effect on stability and purity}

Note from \erf{SBE1} that $y_{\rm c}=0$ is a stationary solution.
Assuming this value for $y_{\rm c}$, we can again use the
polar coordinates $\theta$ and $r$ as in \erf{rtxz}.
Their
equations of motion are found from
\bqa
\theta+d{\theta}=\tan^{-1}\left(\frac{x_{\rm c}+dx_{\rm c}}{z_{\rm
c}+dz_{\rm c}}\right), \\
r+dr=\sqrt{(x_{\rm c}+dx_{\rm c})^2+(z_{\rm c}+dz_{\rm c})^2}.
\eqa
Using the Taylor expansion and \ito rules, and assuming that the state
is initially pure ($r=1$) we find
\beq
dr=0
\eeq
and
\bqa \label{non1}
d{\theta}&=&
\left[2\alpha+2\sqrt{\gamma}\lambda{I(t-\tau)}+\frac{\gamma}{2}{(2+\cos\theta)}{
\sin\theta}\right]dt\nn\\
&&\nl{+} \sqrt{\gamma}(1+\cos\theta)dW(t),
\eqa
where
\beq
I(t-\tau)dt = \sqrt{\gamma}\sin\theta(t-\tau)dt + dW(t-\tau).
\eeq
That is, the state remains pure and
the simulation reduces to the single non-Markovian
stochastic differential equation
(\ref{non1}) which is readily solved in Matlab.

The two plots in Fig.~3 are the typical quantum trajectories in
Bloch space for $\theta_0=\pi/6$, starting at the ground state. The main
plot with more dramatic
fluctuations is the trajectory via the path
through the ground state with feedback time delay
$\tau=0.02\gamma^{-1}$, while the plot with an
inset into this figure is the one for feedback with no time delay and the
trajectory is
via the path through the excited state. Clearly from these two
trajectories, there are two ways for the
system to reach a desired state of $\theta=\theta_{0}$.  As shown
in this figure, the trajectory of the main plot with $\tau=0.02\gamma^{-1}$
continues to evolve
stochastically after the transients have died away. The parameter
$\theta$  wanders around the desired the pure state
$\theta_0=\pi/6$ even when the
system is in steady state. By contrast, that of the inset for Markovian
feedback
stops precisely at the
desired pure steady state with no fluctuations persistent.
Moreover, we find that the longer the
feedback delay time is, the more
dramatic the fluctuations are. Therefore, the longer time delay will result
in more randomness in the steady state quantum trajectory. This indicates
that the
stability of the single trajectory is reduced by the effect of non-zero
feedback delay time.

\begin{figure}[htp]
\center
\centerline{\hbox{
\psfig{figure=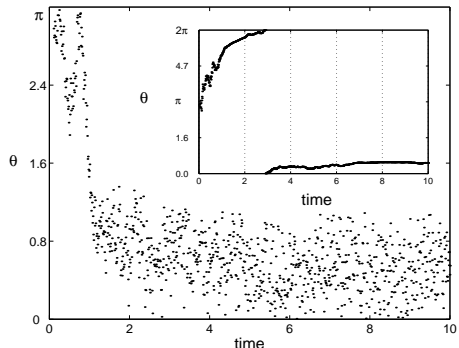,width=6cm}}}
\center
	\caption{ \narrowtext Typical quantum trajectories for $\theta$
	(representing the conditioned state),
for $t\in [0,10\gamma^{-1}]$ and $\theta_0=\pi/6$, starting at the ground
state.
Note that the single trajectory can go either via the ground state or the
excited state
to get to $\theta_0=\pi/6$. The
main plot with a more dramatic fluctuations is for feedback with a large time
$\tau=0.02\gamma^{-1}$, while
the
inset is for feedback with no time delay.}
	\label{fig:delta1}
\end{figure}

Another  typical trajectory, this time for $\theta_0=0$, is shown in
Fig.~4 in Bloch space for
$t\in [0,100\gamma^{-1}]$ with a feedback delay time $\tau=0.02\gamma^{-1}$,
again starting at
the ground state.  Firstly, to consider the
stability of the single trajectory, we see
that the initial evolution is erratic,
then on a time scale of a few $\gamma^{-1}$ the system relaxes towards
the desired
state $\theta_0=0$. As in the previous case,
the fluctuations are still persistent even in steady state and
there is considerable variation around the desired
angle of $\theta_0=0$.
Secondly, to consider the purity of the single trajectory, we can see
from this plot that in a single
trajectory the system state is always pure; it lives on the surface of
the Bloch sphere. However, an ensemble (or time) average of the system
state will put it inside the Bloch sphere. The time-averaged state
from this trajectory (ignoring transients) is represented by the
single star dot
underneath the point $\theta_{0}=0$. The larger the fluctuations
around $\theta_{0}$, the closer this point would be towards the center
of the sphere and the lower the purity of the average state.
 Therefore, we can see that the
purity of a single
trajectory is not be affected by the non-Markovian process while
the purity of the average state is. Rather, the purity of the average state
reflects the stability of the individual trajectory.

\begin{figure}[htp]
\center
\centerline{\hbox{
\psfig{figure=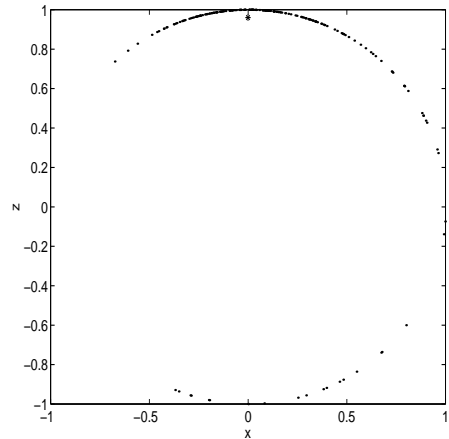,width=6cm}}}
\center
	\caption{\narrowtext Typical single quantum trajectories in Bloch
space for
$t\in [0,100\gamma^{-1}]$ with
$\tau=0.02\gamma^{-1}$ and
$\theta_0=0$, starting at the ground state.
Note that the single star dot underneath the single trajectory path
indicates the loss of
purity of the average state, while the single
trajectory evolves stochastically along the outside full circle which is
the locus of pure states.}
	\label{fig:delta2}
\end{figure}

Fig.~5 is the plot of the locus of the time-averaged states in Bloch
space for three different feedback delay
times $\tau=0, 0.02\gamma^{-1},$ and $0.2\gamma^{-1}$. For each time delay we
plot many points, each corresponding to a different value of
$\theta_{0}$, the angle at which the Markovian feedback would
stabilize the state. We have deliberately symmetrized this plot by
combining the simulations of positive and negative $\theta$.

A number of points are worth noting. First, and most
obviously, the
degree of purity (measured by the $r$, the distance
from the origin)
decreases with $\tau$. Second, the gap at the equator for $\tau=0$
quickly widens for larger delay time $\tau$, so that the purity of the
time-averaged states with
$\theta$ close to $\pi/2$ is decreased, which indicates that the states
near the equator
can not be well-protected against decoherence. Third, the purity of the
time-averaged
states in the upper half of the Bloch sphere is affected much more by
the increase of delay time $\tau$ than those in the lower half. Since
for very long time delay the feedback could not be effective, and
since the
stationary states with no feedback are
confined to the lower half of the Bloch sphere, this is perhaps not
surprising. Fourth, the bottom half curves for the larger delay time
are closer to the ground state even than the no-feedback curve of
Fig.~2. This is due to the fact
that we have not
optimized the driving and feedback amplitude to search out the best
possible pure
state. With a proper choice of driving and feedback amplitude, the
curves should
always lie between the  no-feedback ellipse of Fig.~2 and the $r=1$
pure state circle.

\begin{figure}[htp]
\center
\centerline{\hbox{
\psfig{figure=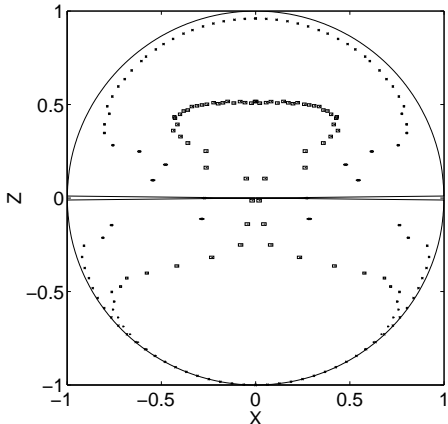,width=6cm}}}
\center
	\caption{\narrowtext Locus of the solutions to the Bloch equations
for feedback with different values of delay time
$\tau$.
 From the outside in, we have $\tau = 0, 0.02\gamma^{-1},$ and
$0.2\gamma^{-1}$.
 The plot with $\tau=0$ is an analytical solution, and the others with
 error bars are
from numerical
simulations. The simulation time for each time-averaged state is
$10^4\gamma^{-1}$, giving an effective sample size of order $10^{4}$,
since $\gamma^{-1}$ is of order the correlation time of the system.}
	\label{fig:delta3}
\end{figure}

All of these features (except the last, which was avoided through
proper optimization) also arose from
 feedback with non-unit-efficiency detection, as
discussed in our previous paper \cite{WangWise}. Both
non-Markovian feedback and  non-unit-efficiency detection decrease the
degree of stability of the system and thus to
limit the capability of feedback decoherence control. The difference
is that with non-unit efficiency detection not only the average, but
also the individual conditioned
state of the system is impure.

\subsection{Exception: the equatorial states }

As mentioned in Sec.~III,
the story for trying to stabilize the equatorial states
$\theta_0=\pm \pi/2$ is special. First, consider the Markovian case.
The SBEs for the equatorial states are
\bqa \left(\begin{array}{c}
dx_{\rm c}\\dy_{\rm c}\\dz_{\rm c}\end{array}\right)
&=& dt \left(\begin{array}{ccc}
0 &0  &  0 \\
0   & -{\gamma}/{2} & 0       \\
0  & 0 &-\gamma/2\\
\end{array}\right)
\left(\begin{array}{c}
x_{\rm c}\\y_{\rm c}\\z_{\rm c}\end{array}\right)
\nl{+}\sqrt{\gamma} dW(t)\left(\begin{array}{c}
1-x_{\rm c}^2\\
-x_{\rm c}y_{\rm c}\\
-x_{\rm c}z_{\rm c}\end{array}\right).
\eqa
Both $z_{c}$ and $y_{\rm c}$ will decay to zero (as required for
$|\theta|=\pi/2$), and their noise terms vanish at that point. By
contrast, the equation for $x_{\rm c}$ is independent of the others, and is
purely stochastic:
\beq
dx_{\rm c} = \sqrt{\gamma}dW(t)(1-x_{\rm c}^{2}).
\eeq

Clearly the equatorial pure states with $x_{\rm c}=\pm 1$ are
stationary solutions to this problem. Also, the system will tend to
one of these states.  But it is also clear that
$x_{\rm c}$ has no preference to go to either of these states. Hence
they are not stable. The ensemble average $x$ is unchanging
under this evolution. Thus a perturbation which moves the state from
$x_{\rm c}=1$ to
$x_{\rm c} = 1-\epsilon$ say, will result in a proportion
$\epsilon/2$ of the states ending up at $x_{\rm c}=-1$, and a
proportion $1-\epsilon/2$ ending up at $x_{\rm c}=1$.

The above discussion all refers to the Markovian case. Having a time
delay will introduce extra noise in the system, as seen in the
preceding section. This is a
sort of perturbation that will disturb the state of the system. Thus
we expect that with a time delay, even if the system is close to one
of the fixed states $x_{\rm c} = \pm 1$, it will not stay there.
Rather, we expect that it will at some point randomly switch to being
close to the
other fixed point. This is indeed what we observe, as we
illustrate  by showing a typical
trajectory for $\theta_0=\frac{\pi}{2}$ in Fig.~6, again starting from
ground state. On a time scale of a few
$\gamma^{-1}$ the system reaches the region of the no-delay fixed
point $x_{\rm c}=1$, on one
side of the equator. Then at a later time, about $5\gamma^{-1}$ here, it
stochastically hops to the other side of the Bloch sphere (going
through the ground state again), ending up near the other fixed
point with $x_{\rm c}=-1$.

\begin{figure}[htp]
\center
\centerline{\hbox{
\psfig{figure=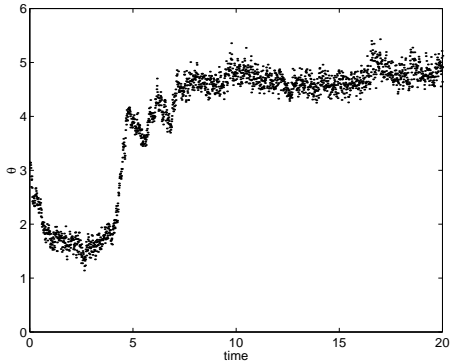,width=6cm}}}
\center
	\caption{\narrowtext  Typical quantum trajectories for
$\theta_0={\pi}/{2}$
with driving amplitude $ \alpha = 0$,
feedback amplitude $\lambda = -\sqrt{\gamma}/{2}$ and
delay time $\tau=0.02\gamma^{-1}$, shown by $\theta$ (dots) as
functions of time $t\in [0,20\gamma^{-1}]$. This plot shows that the system
has no
preference to go to either equatorial point $x_{\rm c}=1$
($\theta={\pi}/{2}$)
or $x_{\rm c}=-1$ ($\theta=3\pi/2$). }
	\label{fig:delta4}
\end{figure}

\subsection{Analytical solution}
As we mentioned in the introduction,
the excited state $\theta_0=0$ is the only state
where we can get both analytical and numerical
solutions to the non-Markovian feedback process. In this subsection, we
will derive
 the analytical solution of the excited state
and compare it with the numerical one.

For short time delay we expect that (except for states near the
equator), $\theta$ will stay near the desired $\theta_{0}$. Thus, it
makes sense to  define $\delta\theta=\theta-\theta_0$ and do a
perturbative expansion for small $\delta \theta$ in  Eq.~(\ref{par1}),
Eq.~(\ref{par2}) and Eq.~(\ref{non1}). This gives the \ito equation
\bqa \label{ana1}
\dot{\delta\theta(t)}&=&-\gamma(1+\cos{\theta_0})
\cos{\theta_0}\delta\theta(t-\tau)\nn\\
&&+\gamma(\cos{\theta_0}+\frac{1}{2}\cos{2\theta_0}){\delta\theta(t)}-
\sqrt{\gamma}\sin\theta_0{\delta\theta}{\xi(t)}\nn\\
&&-\sqrt{\gamma}(1+\cos{\theta_0})\left[\xi(t-\tau)-\xi(t)\right].
\label{ddtdt}
\eqa
Now the purity of the average state can be defined as
\beq
 P=2{\rm Tr}[\rho^{2}]-1 = r^{2}=
 \left<x_{c}\right>^2+\left<y_{c}\right>^2+\left<z_{c}\right>^2. \\
\eeq
Using $y_{c}=0$, $x_{c}=\sin \theta$, $z_{c}=\cos\theta$ and
assuming $\langle\delta\theta\rangle =0$
yields, to leading order,
\beq
P\simeq1-\left<(\delta\theta)^2\right>.
\eeq
Thus if we can find an expression for $\left<(\delta\theta)^2\right>$
we can find the purity of the average state. The decrease in purity
due to fluctuations in $\delta \theta$ can be seen directly in
Fig.~4, for the case $\theta_{0}=0$.

Inspecting \erf{ddtdt} we can see that it is solvable only if
$\sin\theta_{0}=0$ so that the multiplicative term
$\delta\theta\xi(t)$ vanishes. This will be the case only
when one is trying to put the atom  in the excited state $\theta_0=0$, or the
ground state $\theta_{0}=\pi$. The latter is trivial, as it is
achieved exactly with no feedback or driving. The former is the interesting
case, and with $\theta_{0}=0$, Eq.~(\ref{ana1}) becomes
\beq \label{ana2}
\dot{\delta\theta}(t)=-2\gamma\delta\theta(t-\tau)+\frac{3\gamma}{2}{\delta
\theta(t)}-2\sqrt{\gamma}\left[\xi(t-\tau)-\xi(t)\right],\\
\eeq
We show in the appendix that we can solve this equation exactly in
Fourier space, and use the result to evaluate
 $\left<(\delta\theta)^2\right>$ approximately, in the limit
 $\tau \ll \gamma^{-1}$. The result is
\beq \label{ana10}
P=1-4\gamma\tau.
\eeq

\begin{figure}[htp]
\center
\centerline{\hbox{
\psfig{figure=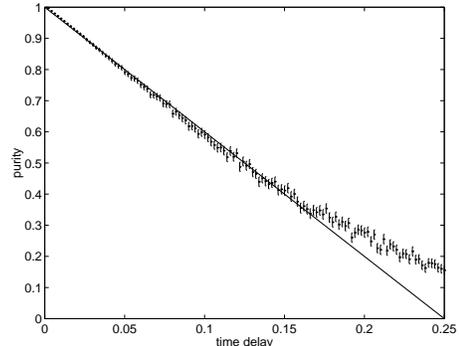,width=6cm}}}
\center
	\caption{\narrowtext Plot of the purity against time delay for the atom
	near the excited state. The solid line is the theoretical result
while the
	dotted line with error bars is the numerical one. The analytical
result agrees
	very well
	with the numerical one when feedback delay time is less than  $
	0.17\gamma^{-1}$. }
	\label{fig:delta5}
\end{figure}

We plot in Fig. 7 the purity against delay time for the atom with
$\theta_{0}=0$, according to the analytical approximation (\ref{ana10})
and from numerical simulations. Clearly the analytical
result agrees very well with the numerical one when the feedback delay time
is less than about
$0.17\gamma^{-1}$.  When the feedback delay time  is larger than
$0.17\gamma^{-1}$, a significant difference appears, and
increases
 as the feedback delay time becomes larger.
However, as we can see from Eq.~(\ref{ana10}), the feedback delay time must
be less than $0.25\gamma^{-1}$ in \erf{ana10} in
order to satisfy $P>0$.
Therefore for the atom near
the excited state, our
analytical solution is a good approximation for a surprisingly large
delay time $\tau$.

\section{Discussion}

In this paper, we have given a rigorous analysis of an anti-decoherence
feedback scheme in a two-level atom in the limit where the feedback time
delay is significant.
We use numerical
simulations to  determine the influence of the non-Markovicity on
the effectiveness of the feedback.
We find that, unlike the case for Markovian feedback,  it is not possible to
stabilize the atom in a
fixed pure state (except the ground state which is trivially pure by
setting the driving and
feedback to zero). Although the conditioned system state always
remains pure (since we are assuming perfect detection) it does not
remain stable but continues to wander around the desired fixed pure
state. Thus the ensemble-averaged state (which is the time-averaged
state in steady state) is not pure, but mixed. As expected,
 the longer the feedback delay time is, the more
dramatic the fluctuations are and consequently the less pure the
average state is.

 We find that the purity of states in the upper
half of the Bloch sphere is affected much more by feedback delay than
those in the lower half. This is not unexpected, as with no feedback
states in upper half cannot be produced at all.
Also, the purity of states near the equator of the Bloch
sphere is very much affected. This is because the feedback algorithm for
stabilizing states on the equator cannot distinguish between
diametrically opposite points on the equator. Because delayed feedback
cannot perfectly cancel the measurement back-action noise, trying to
stabilize at a point on the equator will give a conditioned state which
flips at
random from one side to the other, on a time scale that decreases as
the time delay increases. The unconditioned (or time-averaged)
state is subsequently almost completely mixed. All of these features
are qualitatively similar to the results obtained for no time delay
but with inefficient detection in Ref.~\cite{WangWise}. The major
consequence is that states
which are near-equal superpositions of excited and ground states cannot be
well-protected against decoherence.

There is one case in which it is possible to obtain analytical
results, namely when one tries to stabilize the atom in the excited
state. Our approximate  analytical result that the purity (the square
of the Bloch radius) of the average
state is given by $P=1-4\gamma\tau$ (where $\tau$ is the delay and
$\gamma$ the spontaneous emission rate) agrees very well with our
numerical results even for quite large $\tau$, up to $0.17
\gamma^{-1}$. This shows the usefulness of the quantum trajectory
approach to feedback. Not only is it the only practical way to treat a
non-Markovian problem, it can also yield analytical solutions in some
cases. Other approaches to non-Markovian feedback can only be used
for systems with linear dynamics, and it is not clear that a
linearization procedure would work even for the special case of
stabilizing near the excited state.

The control of decoherence has become an important topic in recent years in
many related  research areas. In quantum computation decoherence is a
main limiting factor. In a quantum computer information is not stored as bits,
but rather as qubits\cite{Nielsen00}. A qubit is a quantum system
with a two dimensional Hilbert space, such as a two-level atom.
The logical basis states are two orthogonal basis states,
$|0\rangle, | 1\rangle$,
but the qubit may be in a superposition of these states. A quantum
computer would consist of a large number $N$ of qubits, and could exist
as an arbitrary state in the $2^{N}$-dimensional Hilbert space.
If any individual qubit undergoes
decoherence,
such as spontaneous emission, the quantum computation is destroyed.
Furthermore this decoherence rate scales linearly with the number of
two-level systems. If quantum computation is to become a reality a means must
be found to control such decoherence.
Sophisticated quantum error correction methods have
been developed to detect and control arbitrary amplitude and phase
errors\cite{Caves98}. While our decoherence-control
scheme is by no means as general as error correction, it may be of use
for maintaining qubits in a particular pure state until required.

We close by saying a few words about experimental realizability.
The results of this work shows that a time delay $\tau$ is not fatal to
controlling decoherence even in a nonlinear system like the two-level
atom. The purity of the state will, in most cases, remain close to
unity as long as $\tau$ is much less than the atomic lifetime
$\gamma^{-1}$. This is feasible with very fast electronics. The
greater difficulty is with obtaining high efficiency in detection.
Collecting a large proportion
of the light emitted by an atom in free space is very
difficult. It is much easier to collect the light emitted from a
cavity, as this propagates in one direction. Therefore, the most
likely scenario for realizing the scheme would be in the context of
cavity QED \cite{Tur98}. If a two-level atom is strongly coupled ($g$)
to a single cavity mode, which is
strongly damped ($\kappa$),
the combined system acts like an effective two-level atom.
The output beam
of the cavity is effectively the spontaneous emission of the atom.
Then, given that
the time delay in the feedback loop is relatively small compared
to the effective atomic lifetime $\sim \kappa/g^{2}$),
we can control  the decoherence
of the cavity QED system.

\section{Acknowledgments}

This work has been supported by the Australian Research Council, the
University of Queensland, and the Department of Employment, Education and
Training, Australia. JW would like to thank Prof. H.
Carmichael
and A/Prof. Z. Ficek for useful discussions.

\section*{Appendix}

For the atom in the excited state, the SBE in the time domain is given by
\bqa
\dot{\delta\theta}(t)&=&-2\gamma\delta\theta(t-\tau)+\frac{3\gamma}{2}{\delta
\theta(t)}\nl{-}2\sqrt{\gamma}\left[\xi(t-\tau)-\xi(t)\right].\label{appSBE}
\eqa
Applying the Fourier transformation, we can solve the SBE
in the frequency domain:
\bqa
\tilde{\delta\theta}(\omega)=\frac{\sqrt{\gamma}(e^{i\omega\tau}-1){\tilde{\xi}
(\omega)}}
{\left[\gamma(\frac{3}{2}-2e^{i\omega\tau})+i\omega\right]}.
\eqa
Applying the inverse Fourier transformation to go back to the time domain,
we get
\beq
{\delta\theta}(t) =\frac{1}{2\pi}\int_{-\infty}^{\infty}d{\omega}
e^{-i{\omega}t}
\frac{2\sqrt{\gamma}(e^{i\omega\tau}-1)}{\gamma(\frac{3}{2}-2e^{i\omega\tau})+i
\omega} \xi(\omega).
\eeq
Since $\left< \xi(\omega)\xi(\omega^{'})\right>=2\pi\delta(\omega+\omega^{'})$,
 we find
\bqa
\left<{\delta\theta}^2(t)\right>&=&
\frac{1}{2\pi}\int_{-\infty}^{\infty}
\frac{4\gamma(2-2\cos{\omega\tau})d{\omega}}
{\gamma^2(\frac{3}{2}-2\cos{\omega\tau})^2+(\omega-2\gamma
\sin{\omega\tau})^2}\nn\\
&=&
\frac{1}{\pi\gamma}\int_{0}^{\infty}
\frac{16\sin^{2}(\omega\tau/{2})d{\omega}}
{(\frac{3}{2}-2\cos{\omega\tau})^2+({\omega}/{\gamma}-2\sin{\omega\tau})^2}.\nn\
\
\eqa

The above integral is too complicated to be useful. Also, we are
really only interested in the limit $\tau \ll \gamma^{-1}$, since that
is the limit in which $\delta\theta$ is expected to be small so that
the approximations leading to \erf{appSBE} will be valid.
With this in mind, consider a parameter $c$ such that
${\gamma} \ll {c} \ll {\tau^{-1}}$, and $c^{3}\tau \ll \gamma^{2}$.
Then split the $\omega$-integration up as
$\int_{0}^{\infty}=\int_{0}^{c}+\int_{c}^{\infty}$. When
${\omega}<c$, we have
$16\sin^{2}{{\omega\tau}/{2}}<(2\omega\tau)^{2}{\ll}1$ in the numerator,
while the denominator
is greater than
or equal to
$\frac{1}{4}$. Therefore the integration from $0$ to $c$ is bounded
above by $16\tau^{2}c^{3}/3$. On the other hand, when $\omega>c$, we
have $\omega\gg \gamma$ so that ${\omega}/{\gamma}$ is much greater than
all the
other terms in the denominator. Therefore we can approximate the
integration in this region as
\beq
\int_{c}^{\infty}d{\omega}{\frac{16\gamma^{2}
\sin^2({\omega\tau}/{2})}{\omega^2}},
\eeq
which, since $c\ll \tau^{-1}$, can be further approximated as
\beq
\int_{0}^{\infty}d{\omega}{\frac{16\gamma^{2}\sin^2{(\omega
\tau}/{2})}{\omega^2}} = \gamma^{2}4\pi\tau
\eeq
Since by assumption $c^{3}\tau \ll \gamma^{2}$, this integral is
always much greater than that from $0$ to $c$. We thus finally arrive
at
\beq
\left<{\delta\theta}^2(t)\right> \simeq
\frac{1}{\pi\gamma}\times \gamma^{2}4\pi\tau = 4\gamma\tau \ll 1.
\eeq

\end{multicols}
\end{document}